\documentclass[12pt,thmsa]{article}
\usepackage{amssymb}
\usepackage{amsfonts}
\usepackage{makeidx}
\usepackage{amsmath}
\usepackage{sw20lart}

\input{tcilatex}

\begin{document}

\author{Ariane Lambert Mogiliansky,\thanks{
PSE, (CNRS, EHESS, ENS, ENPC), Ecole Economique de Paris, alambert@pse.ens.fr%
} Shmuel Zamir,\thanks{%
CREST-LEI Paris and Center for the Study of Rationality, Hebrew University,
zamir@math.huji.ac.il} Herv\'{e} Zwirn\thanks{%
IHPST, UMR 8590 CNRS/Paris 1, and CMLA, UMR 8536 CNRS/ENS Cachan,
herve.zwirn@m4x.org}}
\title{Type Indeterminacy: \\
A Model of the KT(Kahneman--Tversky)-Man\thanks{%
We are grateful for helpful comments from V.I. Danilov, J. Dreze, P. Jehiel,
D. Laibson, P. Milgrom, W. Pesendorfer, A. Roth, and seminar participants at
Harvard, Princeton, and CORE. }}
\date{\today }
\maketitle

\begin{abstract}
In this paper we propose to use elements of the mathematical formalism of
Quantum Mechanics to capture the idea that agents' preferences, in addition
to being typically uncertain, can also be \textit{indeterminate}. They are
determined (i.e., realized, and not merely revealed) only when the action
takes place. An agent is described by a \textit{state}\textbf{\ }that is a
superposition of potential types (or preferences or behaviors). This
superposed state is projected (or \textquotedblleft
collapses\textquotedblright ) onto one of the possible behaviors at the time
of the interaction. In addition to the main goal of modelling uncertainty of
preferences that is not due to lack of information, this formalism seems to
be adequate to describe widely observed phenomena of noncommutativity in
patterns of behavior. We explore some implications of our approach in a
comparison between classical and type indeterminate rational choice
behavior. The potential of the approach is illustrated in two examples.

JEL: D80, C65, B41

Keywords: indeterminacy, non-commutativity, quantum state.

\newpage
\end{abstract}

\section{Introduction}

It has recently been proposed that models of quantum games can be used to
study how the extension of classical moves to quantum ones (i.e., complex
linear combinations of classical moves) can affect the analysis of a game.
For example\ Eisert et al. (1999) show that allowing the players to use
quantum strategies in the Prisoners' Dilemma is a way of escaping the
well-known `bad feature' of this game.\footnote{%
In the classical version of the dilemma, the dominant strategy for both
players is to defect and thereby to do worse than if they had both decided
to cooperate. In the quantum version, there are a couple of quantum
strategies that are both a Nash equilibrium and Pareto optimal and whose
payoff is the one of the joint cooperation.} From a game-theoretical point
of view the approach consists in changing the strategy spaces, and thus the
interest of the results lies in the appeal of these changes.\footnote{%
This approach is closely related to quantum computing. It relies on the use
of a sophisticated apparatus to exploit q-bits' property of entanglement in
mixed strategies.}

This paper also proposes to use elements of the mathematical formalism of
Quantum Mechanics but with a different intention: to model uncertain
preferences.\footnote{%
In this work we borrow the elements of the quantum formalism that concern
the measurement process. We will not use the part of the theory\ theory
concerned with the evolution of systems over time.} The basic idea is that
the Hilbert space model of Quantum Mechanics\textbf{\ }can be thought of as
a very general contextual predictive tool particularly well-suited to
describing experiments in psychology or in \textquotedblleft
revealing\textquotedblright\ preferences.

The well-established Bayesian approach suggested by Harsanyi to model
incomplete information consists of a chance move that selects the types of
the players and informs each player of his own type. For the purposes of
this paper, we underline the following essential implication of this
approach: all uncertainty about a player's type exclusively reflects the
others player's incomplete knowledge of it. This follows from the fact that
a Harsanyi type is fully determined. It is a complete well-defined
description of the characteristics of a player that is known to him.
Consequently, from the point of view of the other players, uncertainty as to
the type can only be due to lack of \textit{information}. Each player has a
probability distribution over the type of the other players, but her own
type is fully determined and is known to her.

This brings us to the first important point at which we depart from the
classical approach: we propose that in addition to informational reasons,
the uncertainty about preferences is due to \textit{indeterminacy}:\textit{\ 
}prior to the moment a player acts, her (behavior) type is indeterminate.
The \textit{state} representing the player is a \textit{superposition} of
potential types. It is only at the moment when the player selects an action
that a specific type is actualized.\footnote{%
The associated concept of irreducible uncertainty, which is the essence of
indeterminacy, is formally defined in Section 2 of the paper.} It is not
merely revealed but rather determined in the sense that prior to the choice,
there is an irreducible multiplicity of potential types. Thus we suggest
that in modelling a decision situation, we do not assume that the preference
characteristics can always be fully known with certainty (neither to the
decision-maker nor even to the analyst). Instead, what can be known is the
state of the agent: a vector in a Hilbert space which encapsulates all
existing information to predict how the agent is expected to behave in
different decision situations.

This idea, daringly imported from Quantum Mechanics to the context of
decision and game theory, is very much in line with Tversky and Simonson
(Kahneman and Tversky 2000) according to whom \textquotedblleft \textit{%
There is a growing body of evidence that supports an alternative conception
according to which preferences are often constructed -- not merely revealed
-- in the elicitation process. These constructions are contingent on the
framing of the problem, the method of elicitation, and the context of the
choice\textquotedblright .} In Ariely, Prelec and Loewenstein (2003), the
authors show in a series of experiments that \textit{\textquotedblright
valuations are initially malleable but become \textquotedblright
imprinted\textquotedblright\ after the agent is called upon to make an
initial decision\textquotedblright } (p. 74). This view is also consistent
with that of cognitive psychology, which teaches one to distinguish between
objective reality and the proximal stimulus to which the observer is
exposed, and to further distinguish between those and the mental
representation of the situation that the observer eventually constructs.
More generally, this view fits in with the observation that players (even
highly rational ones) may act differently in game theoretically equivalent
situations that differ only in seemingly irrelevant aspects (framing, prior
unrelated events, etc.). Our theory as to why agents act differently in game
theoretically equivalent situations is that they are not in the same state;\
(revealed) preferences are contextual because of (intrinsic) indeterminacy.

The basic analogy with Physics, which makes it appealing to adopt the
mathematical formalism of Quantum Mechanics to the social sciences, is the
following: we view decisions and choices as something similar to the result
of a \textit{measurement} (of the player's type). A situation\textbf{\ }of%
\textbf{\ }decision is then similar to an experimental setup to measure the
player's type. It is modelled by\textbf{\ }an operator (called \textit{%
observable}), and the resulting behavior is an eigenvalue of that operator.
The non-commutativity of observables and its consequences (very central
features of Quantum Mechanics) is reminiscent\ of many empirical phenomena
like\ the following one exhibited in a well-known experiment conducted by
Leon Festinger.\footnote{%
Leon Festinger is the father of the theory of cognitive dissonance
(Festinger 1957).} In this experiment people were asked to sort a batch of
spools into lots of twelve and give a square pegs a quarter turn to the
left. They all agreed that the task was very boring. Then,\ they were told
that one subject was missing for the experiment\ and asked to convince a
potential female subject in the waiting room to participate. They were
offered \$1 for expressing their\ enthusiasm for the task.\footnote{%
The experience was richer. People were divided into two groups offered
different amounts of money. For our purpose it is sufficient to focus on a
single result.}\ Some refused, but others accepted. Those who accepted
maintained afterwards that the task was enjoyable. This\ experiment aimed at
showing that attitudes change in response to cognitive dissonance. The
dissonance faced by those who accepted to fake enthusiasm for \$1 was due to
the contradiction\ between the self-image of being "a good guy" and that of\
"being ready to lie for a dollar". Changing one's attitude and persuading
oneself that the task really was interesting was a way to resolve the
dissonance. Similar phenomena have been documented in hazardous industries,
with employees showing very little caution in the face of a\ danger. Here
too, experimental and empirical studies (e.g., Daniel Ben--Horing 1979)
showed how employees\ change their attitude after they have decided\ to work
in a hazardous industry. More generally, suppose that an agent is presented
with the same situation of\ decision in two different contexts. The contexts
may vary with respect to the situation of decision that precedes the
investigated one. In Festinger's experiment the two measurements of the
attitude toward the task differ in that the second one was preceded by a
question about willingness to lie about the task for a dollar. Two contexts
may also vary with respect to the framing of the situation of decision, (cf.
Selten 1998)). If we do not observe the same decision in the two contexts,
then the classical approach considers that the two situations are not
identical and hence that\textbf{\ }they should be modeled so as to
incorporate the context. In many cases, however, such an assumption, i.e.,
that the situations are different, is difficult to justify.

In contrast, we propose that the difference between\textbf{\ }the two\textbf{%
\ }decisions comes from the fact that the agent is not in the same state.
The context, e.g., a past situation of decision to which the agent has been
exposed, is represented\textbf{\ }by an operator that does not commute with
the operator associated with the situation of decision\textbf{\ }currently
considered. The consequence is that\textbf{\ }the initial agent's state has
changed and that the agent is therefore expected to behave differently from
what she would have done if confronted directly with the situation. As in
Quantum Mechanics, the non-commutativity of certain\textbf{\ }situations of%
\textbf{\ }decision (measurements) leads us to conjecture that the
preferences\ of\textbf{\ }an agent are represented by a state that is
indeterminate and gets determined (with respect to any particular type
characteristics) in the course of interaction with the environment. In
section 3, we show how this approach can explain the reversal of preferences
in\textbf{\ }a model of rational choice and that it provides a framework for
explaining cognitive dissonance and framing effects.

The objective of this paper is to propose a theoretical framework for
modelling the KT(Kahneman--Tversky)--man, i.e., for the \textquotedblright
constructive preference perspective\textquotedblright . Our approach amounts
to extending the classical representation of uncertainty in Harsanyi's style
to non-classical indeterminacy. This work is a contribution to Behavioral
Economics.\textbf{\ }A distinguishing feature of behavioral theories is that
they often focus on rather specific anomalies (e.g., `trade-off contrast' or
`extremeness aversion' (Kahneman and Tversky 2000). Important insights have
been obtained by systematically investigating the consequences on utility
maximization of `fairness concerns' (Rabin 1993), `temptation and costly
self-control' (Gul and Pesendorfer 2001) or `concerns for self-image'
(Benabou and Tirole 2002). Yet, other explanations appeal to bounded
rationality, e.g., `superficial reasoning' or `choice of beliefs' (Selten
1998, Akerlof and Dickens 1982). In contrast, the type indeterminacy model
is a framework model that addresses structural properties of preferences,
i.e., their intrinsic indeterminacy. A value of our approach is in providing
a unified explanation for a wide variety of behavioral phenomena.

In section 2, we present the framework and some basic notions of quantum
theory. In Section 3, we develop applications of the theory to social
sciences. In\textbf{\ }Section 4, we discuss some basic assumptions of the
model. The appendix provides a brief exposition of some basic concepts of
Quantum Mechanics.

\section{The basic framework}

In this section we present the basic notions of our framework. They are
heavily inspired by the mathematical formalism of Quantum Mechanics ( see
e.g., Cohen-Tannoudji, Diu, Lalo\"{e} 1973 and Cohen 1989) from which we
also borrow the notation.\ 

\subsection{The notions of state and superposition}

The object of our investigation is the\textbf{\ }individual choice behavior,
which we interpret as the revelation of an agent's preferences in a
situation of decision that we shall call a\textbf{\ }\textit{DS}\textbf{\ (}%
Decision Situation\textbf{).} In this paper, we focus on decision situations
that do not involve any strategic thinking. Examples of such \textit{DS}
include the choice between buying a Toshiba or a Compaq laptop, the choice
between investing in a project or not, the choice between a sure gain of
\$100 or a bet with probability 0.5 to win \$250 and 0.5 to win \$0, etc.
When considering games, we view them as decision situations from the
perspective of a single player who plays once.\footnote{%
All information (beliefs) and strategic considerations are embedded in the
definition of the choices. Thus an agent's play of cooperation in a
Prisoner's Dilemma, is a play of \ cooperation given his information
(knowledge) about the opponent.}

An agent is represented by a state which encapsulates all the\textbf{\ }%
information on the agent's expected choice behavior. The formalism that we
present\textit{\ }below allows for a variety of underlying models. In
particular, some states may correspond to a choice. For instance if the
choice set is $\left\{ a,b,c\right\} ,$\ we may identify three possible
states corresponding to the respective choices \textit{a}, \textit{b} and 
\textit{c}. A state may also\textbf{\ }correspond to an ordering\ of the 3
items in which case we have six possible states (e.g. a state could be
(a,b,c) in this order). We may also consider other models. In section 3.1 we
examine consequences of the second case. For the ease of exposition, the
basic framework is presented in terms of the first case where states are
identified with choices (but the reader is invited to keep in mind that
other interpretations of the model are possible\textbf{)}. However what we
just wrote must be considered in the light of the principle of superposition
as explained below.\textbf{\ }

Mathematically, a state $|\psi \rangle $ is a vector in a Hilbert space $%
\mathcal{H}$ of finite or countably infinite dimensions, over the field of
the real numbers $\mathbb{R}$.\footnote{%
In Quantum Mechanics the field\ that is used is the complex numbers field.
However, for our purposes the field of real numbers provides the structure
and the properties needed (see e.g. Beltrametti and Cassinelli 1981 and
Holland 1995). Everything we present in the appendix (Elements of quantum
mechanics) remains true when we replace Hermitian operators with real
symmetric operators.} The relationship between $\mathcal{H}$ and a decision
situation will be specified later. For technical reasons related to the
probabilistic content of the state, each state vector has to be of length
one, that is, $\langle \psi \left\vert \psi \right\rangle ^{2}=1\ $(where $%
\langle \cdot \left\vert \cdot \right\rangle $ denotes the inner product in $%
\mathcal{H)}$. So all vectors of the form $\lambda \left\vert \psi
\right\rangle $ where $\lambda \in \mathbb{R},$\ correspond to the same
state, which we represent by a vector of length one.

A key ingredient in the formalism of indeterminacy is \textit{the principle
of superposition.} This principle states that the linear combination of any
two states$\;$is itself a possible state.\footnote{%
We use the term state to refer to `pure state'. Some people use the term
state to refer to mixture of pure states. A mixture of pure states combines
indeterminacy with elements of incomplete information. They are represented
by the\textbf{\ }so called density operators.} Consider two states $%
\left\vert \varphi _{1}\right\rangle ,\left\vert \varphi _{2}\right\rangle
\in \mathcal{H}$. If\textbf{\ }$\left\vert \psi \right\rangle =\lambda
_{1}\left\vert \varphi _{1}\right\rangle +\lambda _{2}\left\vert \varphi
_{2}\right\rangle $ with $\lambda _{1},\lambda _{2}\in \mathbb{R}\ $then%
\textbf{\ }$\left\vert \psi \right\rangle \in \mathcal{H}.$\ The principle
of superposition implies that, unlike the Harsanyi type space, the state
space is non-Boolean.\footnote{%
The distributivity condition defining a Boolean space is dropped for a
weaker condition called ortho-modularity. The basic structure of the state
space is that of a logic, i.e., an orthomodular lattice. For a good
presentation of Quantum Logic, a concept introduced by Birkhoff and Von
Neuman (1936), and further developed by Mackey (2004, 1963), see Cohen
(1989).}

\subsection{The notions of measurement\textit{\ }and of observable}

Measurement is a central notion in our framework\textbf{.} A measurement is
an operation (or an\textbf{\ }experiment) performed on a system. It yields a
result, the outcome of the measurement. A defining feature of a measurement
is the so-called first-kindness property.\footnote{%
The term first-kind measurement was introduced by Pauli.} It refers to the
fact that if one performs a measurement on a system and obtain\textbf{s} a
result, then one will get the same result if one\textbf{\ }performs again
that measurement on the same system immediately after. Thus, the outcome of
a first-kind measurement is reproducible but only in a next subsequent
measurement. First-kindness does not entail that the first outcome is
obtained when repeating a measurement if other measurements are performed on
the system in between. Whether an operation is a measurement, i.e., is
endowed with the property of first-kindness is an empirical issue. When it
comes to decision theory, it means that we do not, a priory, assume that any
choice set can be used to measure preferences. In particular, the product
set of two sets each of which is associated with a first-kind measurement,
is not in general associated with a first-kind measurement.\textbf{\ }The
set of decision problems that we consider consists exclusively of decision
problems that can be associated with first-kind measurements. We call those
decision problems Decision Situations\ (\textit{DS}).\footnote{%
Even standard decision theory implicitely restricts its application to
decision problems satisfying the first-kindness property (or that can be
derived from such decision problems). In contrast, random utility models do
not require choice behavior to satisfy the first-kindness property in the
formulation used in this paper.}

A Decision Situation $A$\ can be thought of as an experimental setup where
the agent is invited to choose a particular action among all the\textbf{\ }%
possible actions\textbf{\ }allowed by this Decision Situation. In this
paper, we will consider only the case of finitely many possible outcomes.
They will be labelled from 1 to $n$ by convention. When an agent selects an
action\textbf{,} we say that she `plays' the \textit{DS}. To every Decision
Situation $A$, we will\textbf{\ }associate an \textit{observable,} namely, a
specific symmetric operator on $\mathcal{H}$ which, for notational
simplicity, we also denote by $A$.\footnote{%
Observables in Physics are represented by Hermetian operators because QM is
defined over the field of complex numbers. Here, we confine ourselves to the
field of real numbers which is why observables are represented by symmetric
operators.} If we consider\textbf{\ }only one\textbf{\ }Decision Situation $%
A $ with $n$ possible outcomes, we can assume that the associated Hilbert
space is $n$-dimensional and\textbf{\ }that the eigenvectors of the
corresponding observable, which we denote by $|1_{A}\rangle ,|2_{A}\rangle
,...,|n_{A}\rangle ,$\ all correspond to different eigenvalues, denoted by $%
1_{A},2_{A},...,n_{A}$ respectively. By convention, the eigenvalue $i_{A}$
will be associated with choice $i.$\textbf{\ } 
\begin{equation*}
A\left\vert k_{A}\right\rangle =k_{A}\left\vert k_{A}\right\rangle
,\;\;k=1,...,n.
\end{equation*}%
\ As $A\;$is\ symmetric, there is a unique orthonormal basis of the relevant
Hilbert space $\mathcal{H\;}$formed with its eigenvectors. The basis $%
\left\{ |1_{A}\rangle ,|2_{A}\rangle ,...,|n_{A}\rangle \right\} $ is the
unique orthonormal basis of $\mathcal{H}$ consisting of eigenvectors of $A.$
It is thus possible to represent the agent's state as a superposition of
vectors of this basis:

\begin{equation}
\left| \psi \right\rangle \ =\underset{k=1}{\overset{n}{\sum }}\lambda
_{k}\left| k_{A}\right\rangle ,  \label{exp}
\end{equation}
where $\lambda _{k}\in \mathcal{\mathbb{R}},\forall k\in \{1,...,n\}$ and $%
\underset{k=1}{\overset{n}{\sum }}\lambda _{k}^{2}=1$.

The Hilbert space can be decomposed as follows 
\begin{equation}
\mathcal{H=H}_{1_{A}}\mathcal{\oplus },...,\mathcal{\oplus H}_{n_{A}},\;%
\mathcal{H}_{i_{A}}\perp \mathcal{H}_{j_{A}},\;i\neq j,  \label{exp2}
\end{equation}%
where $\oplus $ denotes the direct sum of the subspaces $\mathcal{H}%
_{1_{A}},...,\mathcal{H}_{n_{A}}$ spanned by $|1_{A}\rangle
,...,|n_{A}\rangle $ respectively.\footnote{%
That is, for $i\neq j\ $any vector in $\mathcal{H}_{i_{A}}$ is orthogonal to
any vector in$\;\mathcal{H}_{j_{A}}$ and any vector in $\mathcal{H}$ is a
sum of $n$ vectors, one in each component space.} Or, equivalently, we can
write $I_{\mathcal{H}}=\;P_{1_{A}}+,...,+P_{n_{A}}$ where $P_{i_{A}}$ is the
projection operator on $\mathcal{H}_{i_{A}}$ and $I_{\mathcal{H}}$ is the
identity operator on $\mathcal{H}.$\bigskip

A Decision Situation $A$\ is an experimental setup and the actual
implementation of the experiment is represented by a \textit{measurement} of
the associated observable $A$. According to the so-called Reduction Principle%
\textit{\ }(see Appendix), the result of such a measurement can only be one
of the $n$ eigenvalues of $A$. If the result is $m_{A},$ i.e., the player
selects action $m,$ the superposition $\sum \lambda _{i}\left\vert
i_{A}\right\rangle $ \textquotedblleft collapses\textquotedblright\ onto the
eigenvector associated with the eigenvalue $m_{A}.$ The initial state $%
\left\vert \psi \right\rangle $ is projected into the subspace $\mathcal{H}%
_{m_{A}}$(of eigenvectors of $A$ with eigenvalue $m_{A}).$ The probability
that the measurement yields the result $m_{A}$ is\ equal to $\langle
m_{A}\left\vert \psi \right\rangle ^{2}=\lambda _{m}^{2},$ i.e., the square
of the corresponding coefficient in the superposition. The coefficients
themselves are called `amplitudes of probability'. They play a key role when
studying sequences of measurements (see Section 2.3). As usual, we interpret
the probability of $m_{A}$ either as the probability that one agent in state 
$\left\vert \psi \right\rangle $ selects action $m_{A}$ or as the proportion
of the agents who will make the choice $m_{A}$ in a population of many
agents, all in the state $\left\vert \psi \right\rangle $.

\ 

In our theory an agent is represented by a state. We shall also use the term
type (and eigentype) to denote a state corresponding\textbf{\ }to one
eigenvector, say $\left\vert m_{A}\right\rangle .\;$An agent in this state
is said to be of type $m_{A}.$\ An agent in a general state $\left\vert \psi
\right\rangle $\ can be expressed as a superposition of all eigentypes of
the \textit{DS} under consideration.$\ $Our notion of type is closely
related to the notion introduced by Harsanyi. Consider a simple choice
situation e.g., when an employee faces a menu of contracts. The type
captures all the agent's characteristics (taste, subjective beliefs, private
information) of relevance for uniquely predicting the agent's behavior. In
contrast to Harsanyi, we shall not assume that there exists an exhaustive
description of the agent that enables us to determine the agent's choice
uniquely and \textit{simultaneously} in all possible Decision Situations\ .
Instead, our types are characterized by an irreducible uncertainty that is
revealed when the agent is confronted with a sequence of \textit{DS} (see
Section 2.3.2 below for a formal characterization of irreducible
uncertainty).

\ 

\textit{Remark: }Clearly, when only one \textit{DS} is considered, the above
description is equivalent to the traditional probabilistic representation of
an agent by a probability vector $(\alpha _{1,....,}\alpha _{n})\;$in which $%
\alpha _{k}$ is the probability that the agent will choose action $k_{A}$
and $\alpha _{k}=\lambda _{k}^{2}$ for $k=1,...,n.$ The advantage of the
proposed formalism consists in enabling us to study several decision
situations and the interaction between them.

\subsection{More than one Decision Situation}

When studying more than one \textit{DS}, say $A$ and $B$, the key question
is whether the corresponding observables are commuting operators in $%
\mathcal{H},$ i.e., whether $AB=BA.$\ Whether two \textit{DS}\ can be
represented by two commuting operators or not is an empirical issue. We next
study its mathematical implications.

\subsubsection{Commuting Decision Situations}

Let $A$ and $B$ be two \textit{DS}. If the corresponding observables commute
then there is an orthonormal basis of the relevant Hilbert space $\mathcal{H}
$ formed by eigenvectors common to both $A$ and $B$. Denote by $\left\vert
i\right\rangle $ (for $i=1,...,n)\;$these basis vectors.\ We have

\begin{equation*}
A\left\vert i\right\rangle =i_{A}\left\vert i\right\rangle \;\text{and }%
B\left\vert i\right\rangle =i_{B}\left\vert i\right\rangle .
\end{equation*}%
In general, the eigenvalues can be degenerated (i.e., for some $i$ and\ $j,$ 
$_{{}}i_{A}=j_{A}$ or $i_{B}=j_{B}).$\footnote{%
In the argument that we develop in section 3.1, the pure states are linear
orders therefore choice experiments are observables with degenerated
eigenvalues} Any normalized vector $\left\vert \psi \right\rangle $ of $%
\mathcal{H}$ can be written in this basis:

\begin{equation*}
\left\vert \psi \right\rangle =\sum_{i}\lambda _{i}\left\vert i\right\rangle
,
\end{equation*}
where $\lambda _{i}\in \mathcal{\mathbb{R}}$, and $\sum_{i}\lambda
_{i}^{2}=1.\;$If we measure $A$ first, we observe eigenvalue $i_{A}$ with
probability

\begin{equation}
p_{A}\left( i_{A}\right) =\sum_{j;j_{A}=i_{A}}\lambda _{j}^{2}.  \label{CO1}
\end{equation}%
\ If we measure $B$ first, we observe eigenvalue $j_{B}$ with probability $%
p_{B}\left( j_{B}\right) =\sum_{k;k_{B}=j_{B}}\lambda _{k}^{2}.\;$After $B$
is measured and the result $j_{B}$ is obtained$,$ the state $\left\vert \psi
\right\rangle $ is projected into\ the eigensubspace $\mathcal{E}_{j_{B}}$
spanned by the eigenvectors of $B$\ associated with $j_{B}.\;$More
specifically, it collapses onto the state:

\begin{equation*}
\left\vert \psi _{j_{B}}\right\rangle =\frac{1}{\sqrt{\sum_{k;k_{B}=j_{B}}%
\lambda _{k}^{2}}}\sum_{k;k_{B}=j_{B}}\lambda _{k}^{{}}\left\vert
k\right\rangle
\end{equation*}
(the factor $\frac{1}{\sqrt{\sum_{k;k_{B}=j_{B}}\lambda _{k}^{2}}}$ is
necessary to make $\left\vert \psi _{j_{B}}\right\rangle $ a unit vector).

When we measure $A$ on the agent in the state $\left\vert \psi
_{j_{B}}\right\rangle ,$ we obtain $i_{A}$ with probability 
\begin{equation*}
p_{A}\left( i_{A}|j_{B}\right) =\frac{1}{\sum_{k;k_{B}=j_{B}}\lambda _{k}^{2}%
}\sum_{\substack{ k;k_{B}=j_{B}  \\ \text{and\ }k_{A}=i_{A}}}\lambda
_{k}^{2}.
\end{equation*}
So when we measure first $B$ and then $A,$ the probability of observing the
eigenvalue $i_{A}$ is $p_{AB}\left( i_{A}\right) =\sum_{j}p_{B}\left(
j_{B}\right) p_{A}\left( i_{A}|j_{B}\right) $:

\begin{equation*}
\begin{array}{lll}
p_{AB}\left( i_{A}\right) & = & \sum_{j}\frac{1}{\sum_{k;k_{B}=j_{B}}\left%
\vert \lambda _{k}\right\vert ^{2}}\sum_{k;k_{B}=j_{B}}\lambda _{k}^{2}\sum 
_{\substack{ _{_{\substack{ l;l_{B}=j_{B}\;  \\ \text{and }l_{A}=i_{A}}}} 
\\ }}\lambda _{l}^{2} \\ 
& = & \sum_{j}\sum_{\substack{ _{_{\substack{ l;l_{B}=j_{B}\;  \\ \text{and\ 
}l_{A}=i_{A}}}}  \\ }}\lambda _{l}^{2}=\sum_{\substack{ _{_{l;l_{A}=j_{A}%
\;}}  \\ }}\lambda _{l}^{2}.%
\end{array}%
\end{equation*}
Hence, $p_{AB}\left( i_{A}\right) =p_{A}\left( i_{A}\right) ,$ $\forall i,\;$%
and similarly $p_{BA}\left( j_{B}\right) =p_{B}\left( j_{B}\right) ,\forall
j.\;$

\smallskip

When dealing with commuting observables it is meaningful to speak of
measuring them\ simultaneously. Whether we measure first $A$ and then $B$ or
first $B$ and then $A$, the probability distribution on the joint outcome is 
$p\left( i_{A}\wedge \text{ }j_{B}\right) =\sum_{_{\substack{ k;k_{B}=j_{B} 
\\ \text{and}\;k_{A}=i_{A}}}}\lambda _{k}^{2},$\ so$\;\left( i_{A\text{ }}%
\text{, }j_{B}\right) $ is a well-defined event.\ Formally, this implies
that the two \textit{DS} can be merged into a single \textit{DS. }When we
measure it, we obtain a vector as the outcome, i.e., a value in $A\ $and a
value in $B$. To each eigenvalue of the merged observable we associate a
type that captures all the characteristics of the agent relevant to her
choices (one in each \textit{DS}).

\textit{Remark: }Note that, as in the case of a single \textit{DS}, for two
such commuting \textit{DS} our model is equivalent to a standard (discrete)
probability space in which the elementary events are $\left\{ \left(
i_{A},j_{B}\right) \right\} $ and$\;p\left( i_{A}\wedge \text{ }j_{B}\right)
=\sum_{\substack{ k;k_{B}=j_{B}  \\ \text{and}\;k_{A}=i_{A}}}\lambda
_{k}^{2}. $ In particular, in accordance with the calculus of probability we
see that the conditional probability\ formula\ holds: 
\begin{equation*}
p_{AB}(i_{A}\wedge j_{B})=p_{A}(i_{A})\ p_{B}(j_{B}|i_{A}).
\end{equation*}%
This also means that the type space associated with type characteristics
represented by commuting observables is equivalent to the Harsanyi type
space. In particular, if the Decision Situations $A$\ and $B$\ together
provide a full characterization of the agent, then all types $i_{A}j_{B}$\
are mutually exclusive:\ knowing that the agent is of type $1_{A}2_{B}$\ it
is sure that she is \textit{not} of type $i_{A}j_{B}$\ for $i\neq 1$\ and/or 
$j\neq 2$.

\ 

As an example, consider the following two Decision Situations. Let $A$ be
the Decision Situation presenting a choice between a week vacation in
Tunisia and a week vacation in Italy. And let $B\;$be the choice between
buying 1000 euros of shares of\textbf{\ }Bouygues Telecom or of Deutsche
Telecom. It is quite plausible that $A$ and $B$ commute, but whether or not
this is the case is of course an empirical question.\ If $A$ and $B$
commute,\ we expect that\textbf{\ }a decision on portfolio ($B$) will\textbf{%
\ }not affect the decision-making concerning the vacation ($A)$. And thus
the order in which the decisions are made does not matter as in the
classical model.

\ 

Note finally that the commutativity of the observables does not exclude
statistical correlations between observations. To see this, consider the
following example in which $A$ and $B\;$each have two degenerated
eigenvalues in a four dimensional Hilbert space. Denote by $\left\vert
i_{A}j_{B}\right\rangle \;(i=1,2,\;j=1,2)\;$the eigenvector associated with
eigenvalues $i_{A}$ of $A$ and $j_{B}$ of $B,\;$and\ let\ the state $%
\left\vert \psi \right\rangle $ be given by

$\left| \psi \right\rangle =\sqrt{\frac{3}{8}}\left| 1_{A}1_{B}\right\rangle
+\sqrt{\frac{1}{8}}\left| 1_{A}2_{B}\right\rangle +\sqrt{\frac{1}{8}}\left|
2_{A}1_{B}\right\rangle +\sqrt{\frac{3}{8}}\left| 2_{A}2_{B}\right\rangle $

Then,$\ \ p_{A}\left( 1_{A}|1_{B}\right) =\frac{\frac{3}{8}}{\frac{3}{8}+%
\frac{1}{8}}=\frac{3}{4},\ \ p_{A}\left( 2_{A}|1_{B}\right) =\frac{\frac{1}{8%
}}{\frac{3}{8}+\frac{1}{8}}=\frac{1}{4}.$

So if we first measure $B$ and find, say, $1_{B},$\ it is much more likely
(with probability $\frac{3}{4}$) that when measuring $A$ we will find $1_{A}$
rather than $2_{A}\;$(with probability $\frac{1}{4}$)$.$ But the two \textit{%
interactions} (measurements) do not affect each other, i.e., the
distribution of the outcomes of the measurement of $A\;$is the same whether
or not we measure $B$ first.

\subsubsection{Non-commuting Decision Situations}

It is when we consider Decision Situations associated with observables that
do not commute that the predictions of our model differ from those of the
probabilistic one. In such a context, the quantum probability calculus ($%
p(i_{A}\left\vert \psi \right. )=\left\langle i_{A}\right. \left\vert \psi
\right\rangle ^{2}$)$\;$generates cross-terms also called interference
terms. These cross-terms are the signature of indeterminacy. In the next
section, we demonstrate how this feature can capture the phenomenon of
cognitive dissonance as well as that of framing.

\ 

Consider two Decision Situations $A$\ and $B$\ with the same number $n$\ of
possible choices. We shall assume for simplicity that the corresponding
observables $A$\ and $B$\ have non-degenerated eigenvalues $1_{A},2_{A},...,$%
\ $n_{A}$\ and $1_{B},2_{B},...,$\ $n_{B}$\ respectively. Each set of
eigenvectors $\left\{ |1_{A}\rangle ,|2_{A}\rangle ,...,|n_{A}\rangle
\right\} $\ and $\left\{ |1_{B}\rangle ,|2_{B}\rangle ,...,|n_{B}\rangle
\right\} $\ is an orthonormal basis of the relevant Hilbert space. Let $%
\left\vert \psi \right\rangle $\ be the initial state of the agent 
\begin{equation}
\left\vert \psi \right\rangle \ =\underset{i=1}{\overset{n}{\sum }}\lambda
_{i}|i_{A}\rangle =\overset{n}{\underset{j=1}{\sum }}\nu _{j}\left\vert
j_{B}\right\rangle .  \label{statevector}
\end{equation}

We note that each set of eigenvectors of the respective observables forms a
basis of the state space. The multiplicity of alternative basis is a
distinguishing feature of this formalism. It implies that there is no single
or privileged way to describe (express) the type of the agent. Instead,
there exists a multiplicity of equally informative alternative ways to
characterize the agent.\ 

We shall now compute the probability for type $i_{A}$\ under two different
scenarios. In the first scenario we measure $A$\ on the agent in state $%
\left\vert \psi \right\rangle .$\ In the second scenario we first measure $B$%
\ on the agent in state $\left\vert \psi \right\rangle $\ and thereafter
measure $A$\ on the agent in the state resulting from the first
measurement.\ We can write observable $B^{\prime }s$\ eigenvector $%
|j_{B}\rangle \ $in the basis made of $A^{\prime }$s eigenvectors: 
\begin{equation}
|j_{B}\rangle =\underset{i=1}{\overset{n}{\sum }}\mu _{ij}|i_{A}\rangle .
\label{Bvector}
\end{equation}%
Using the expression above we write the last term in equation (\ref%
{statevector}) 
\begin{equation}
\left\vert \psi \right\rangle \ =\overset{n}{\underset{j=1}{\sum }}\underset{%
i=1}{\overset{n}{\sum }}\nu _{j}\mu _{ij}|i_{A}\rangle .  \label{state2}
\end{equation}%
From expression (\ref{state2}), we derive the probability $p_{A}(i_{A})\ $%
that the agent chooses $i_{A}$\ $\ $in the first scenario:\textbf{\ }$%
p_{A}(i_{A})=\left( \underset{j=1}{\overset{n}{\sum }}\nu _{j}\mu
_{ij}^{{}}\right) ^{2}$\textbf{. }In the second scenario, she first plays$%
\;B.$\ By (\ref{statevector}) we see that she selects action $j_{B}$\ with
probability $\nu _{j}^{2}.$\ The state $\left\vert \psi \right\rangle \ $is
then projected onto $|j_{B}\rangle .$\ When the state is $|j_{B}\rangle ,\ $%
the\ probability for $i_{A}$\ is given by (\ref{Bvector}),\ it is $\mu
_{ij}^{2}.$\ Summing up the conditionals, we obtain the (ex-ante)
probability for $i_{A}$\ when the agent first plays $B$\ and then $A:$\ $%
p_{AB}(i_{A})=$\ $\overset{n}{\underset{j=1}{\sum }}\nu _{j}^{2}\mu
_{ij}^{2},$\ which is, in general, different from $p_{A}(i_{A})=\left( 
\overset{n}{\underset{j=1}{\sum }}\nu _{j}\mu _{ij}\right) ^{2}.$\ Playing
first $B$ changes the way $A$\ is played. The difference stems from the
so-called interference terms

\begin{eqnarray*}
p_{A}(i_{A}) &=&\left( \overset{n}{\underset{j=1}{\sum }}\nu _{j}\mu
_{ij}\right) ^{2}=\overset{n}{\underset{j=1}{\sum }}\nu _{j}^{2}\mu
_{ij}^{2}+\underset{\text{Interference term}}{\underbrace{2\underset{j\neq
j^{\prime }}{\sum }\left[ \left( \nu _{j^{\prime }}\mu _{ij}\right) \left(
\nu _{j}^{{}}\mu _{ij^{\prime }}^{{}}\right) \right] \qquad }} \\
&=&p_{AB}(i_{A})+\text{interference term}
\end{eqnarray*}
The interference term is the sum of cross-terms involving the amplitudes of
probability (the Appendix provides a description of interference effects in
Physics)\textbf{.}

Some intuition about interference effects may be provided using the concept
of "propensity" due to Popper (1992). Imagine an agent's mind as a system of
propensities to act (corresponding to different possible actions). As long
as the agent is not required to choose an action in a given\ \textit{DS},
the corresponding propensities coexist in her mind; the agent has not
\textquotedblleft made up her mind\textquotedblright . A decision situation
operates on this state of "hesitation" to trigger the emergence of a single
type (of behavior). But as long as alternative propensities are present in
the agent's mind, they affect choice behavior by increasing or decreasing
the probability of the choices in the \textit{DS} under investigation.

\ 

An illustration of this kind of situation may be supplied by the experiment
reported in Knetz and Camerer (2000). The two studied \textit{DS} are the
Weak Link (WL) game and the Prisoners' Dilemma (PD) game.\footnote{%
The Weak Link game is a type of coordination game where each player picks an
action from a set of integers. The payoffs are defined in such a manner that
each player wants to select the minimum of the other players but everyone
wants that minimum to be as high as possible.} They compare the distribution
of choices in the Prisoners' Dilemma (PD) game when it is preceded by a Weak
Link (WL) game and when only the PD game is being played. Their results show
that playing the WL game affects the play of individuals in the PD game. The
authors appeal to an informational argument, which they call the
\textquotedblleft precedent effect\textquotedblright .\footnote{%
The precedent effect hypothesis is as follows: \textquotedblleft The shared
experience of playing the efficient equilibrium in the WL game creates a
precedent of efficient play strong enough to (...) lead to cooperation in a
finitely repeated PD game\textquotedblright ,\ Knetz and Camerer (2000 see
p.206).} However, they cannot explain the high rate of cooperation (37.5 \%)
in the last round of the PD game (Table 5, p. 206). Instead, we propose that
the WL and the PD are two \textit{DS} that do not commute. In such a case we
expect a difference in the distributions of choices in the (last round of
the) PD depending on whether or not it was preceded by a play of the WL or
another PD game. This is because the type of the agent is being modified by
the play of the WL game. \ 

\ 

\textit{Remark: }In the case where\textbf{\ } $A$ and $B$ do not commute,
they cannot have simultaneously defined values: the state of the agent is
characterized by an\textbf{\ }\textit{irreducible uncertainty}. Therefore,
and in contrast with the commuting case, two non-commuting observables
cannot be merged into one single observable. There is no probability
distribution on the events of the type "to have the value $i_{A}$ for $A$
and the value $j_{B}$ for $B"$. The conditional probability formula does not
hold:

\begin{equation*}
p_{A}(i_{A})\neq \overset{n}{\underset{j=1}{\sum }}p_{B}(j_{B})\
p(i_{A}|j_{B}).
\end{equation*}

Indeed, $\left( \overset{n}{\underset{j=1}{\sum }}\nu _{j}\mu _{ij}\right)
^{2}=p_{A}(i_{A})\neq \overset{n}{\underset{j=1}{\sum }}p_{B}(j_{B})\
p(i_{A}|j_{B})=\overset{n}{\underset{j=1}{\sum }}\nu _{j}^{2}\mu _{ij}^{2}.\ 
$

Consequently,\ the choice experiment consisting of asking the agent to
select a pair $\left( i_{A},j_{B}\right) $\ out the set of alternatives $%
A\times B$\ is NOT a \textit{DS.}

We must here acknowledge a fundamental distinction between the type space of
our TI-model (Type Indeterminacy model) and that of Harsanyi. In the
Harsanyi type space the (pure) types are all mutually exclusive: the agent
is either of type $\theta _{i}$\ \textit{or} of type $\theta _{j}$\ but she
cannot be both!$\ $In the TI-model this is not always the case. When dealing
with (complete) non-commuting $\mathit{DS}$\footnote{%
We say that a DS is complete when its outcome provides a complete
characterization of the agent.}, the types associated with respective $%
\mathit{DS}\ $are not mutually exclusive: knowing that the agent is of type $%
1_{A},$\ which is a full description of her type, it cannot be said that she
is \textit{not} of type $1_{B}.\ $The eigentypes of non-commuting \textit{DS}
are \textquotedblleft connected\textquotedblright\ in the sense that the
agent can transit from one type to another under the impact of a
measurement. When making a measurement of $B$\ on the agent of type $1_{A},$%
\ she is projected onto one of the eigenvectors of $\mathit{DS}\ B.$\ Her
type changes from being $1_{A}$\ to being some $j_{B}.$\ 

\section{The Type Indeterminacy model and Social Sciences}

The theory of choice exposed in this paper does not allow for a
straightforward comparison with standard choice theory. Significant further
elaboration is required. Yet, some implications of the type indeterminacy
approach can be explored. First, we shall be interested in comparing the
behavior of an agent of indeterminate type with the one of a classical agent
in the case where both satisfy the Weak Axiom of Revealed Preference (WARP)%
\footnote{%
Samuelson (1947)}\textbf{\ }which is a basic axiom of rational choice. Then,
we will show that this framework can be used to explain two instances of
behavioral anomalies that have been extensively studied in the literature.

\subsection{The TI-model and the classical rational man}

In standard decision theory, it is assumed that an individual has
preferences (i.e. a complete ranking or a complete linear ordering) on the
universal set of alternatives $X.\ $The individual knows her preferences
while the outsiders may not know them. But it is also possible that the
individual only knows what she would choose from some limited sets of
alternatives and not from the whole set $X$. Thus, a less demanding point of
view consists in representing the choice behavior by a choice structure
(i.e. a family $\mathcal{B}$ of subsets of the universal set of alternatives 
$X$ \ and a choice rule $C$ that assigns a non empty set of chosen elements $%
C(A)$ for all $A\in \mathcal{B}$). The link between the two points of view
is well-known. From preferences, it is always possible to build a choice
structure but the reverse is not always true. For it to be true, the choice
structure must display a certain amount of consistency (satisfying the Weak
Axiom of Revealed Preference) and the family $\mathcal{B}$ must include all
subsets of $X$ of up to three elements\footnote{%
See for example Mas-Colell, Whinston and Green, Microeconomic Theory, p.13.}%
. So, preferences can be revealed by asking the individual to make several
choices from subsets of $X.\ $How does this simple scheme changes when we
are dealing with an individual whose type is indeterminate?

\smallskip

Consider a situation where an individual is invited to make a choice of one
item out of a set $A$, $A\subseteq X.$\ If this experiment satisfies the
first-kindness property, we can consider it to be a measurement represented
by an observable. The set of possible outcomes of this experiment is the set 
$A.$\thinspace $\ $We also denote the observable by $A$.\footnote{%
The use of the same symbol for sets of items and observables should not
confuse the reader. Either the context unambiguously points to the right
interpretation or we make it precise.}

\ 

We make two key assumptions on the individual choice behavior:

A1. Choices out of a \textquotedblleft small\textquotedblright\ subset\
satisfy the first-kindness property (the meaning of \textquotedblright
small\textquotedblright\ will be made precise).

A2. Choices out of a \textquotedblleft small\textquotedblright\ subset
respect our Weak Axiom of Revealed Preference (WARP', see below).

\ 

Assumption A1 means that \textquotedblright small\textquotedblright\ subsets
are associated with \textit{DS,} i.e., the experiment consisting in letting
an individual choose an item out of a \textquotedblright
small\textquotedblright\ subset of item is a measurement. Assumption A2
means that choices from \textquotedblleft small\textquotedblright\ subsets
are rational. The idea behind these assumptions is that an individual can,
in her mind, structure any \textquotedblright small\textquotedblright\ set
of alternatives, i.e., simultaneously compare those alternatives. She may
not be able to do that within a \textquotedblleft big\textquotedblright\ set
though. But\textbf{\ }this does not mean that our individual cannot make a
choice from a \textquotedblleft big\textquotedblright\ set. For example, she
might use an appropriate sequence of binary comparisons and select the last
winning alternative. However, such a compound operator would not in general
satisfy the first-kindness property i.e., there may not exist any \textit{DS 
}representing such an operation.

\medskip

A standard formulation of the WARP can be found in Mas-Colell et al. (1995).%
\footnote{%
"If for some $B\in \mathcal{B}\ $with $x,\ y\in B$ we have \thinspace $x\in
C\left( B\right) ,$ then for any $B^{\prime }\in \mathcal{B}$\thinspace $\ $%
with$\ x,y\in B^{\prime }$ and $y\in C\left( B^{\prime }\right) ,$ we must
also have $x\in C\left( B^{\prime }\right) ."($Mas-Colell et al. (1995)
p.10).} We shall use a stronger version by assuming that $C\left( B\right)
,\ $for any $B\in \mathcal{B},\mathcal{\ }$is a singleton.\footnote{%
At this stage of the research we do not want to deal with indifference
relations.} For our purpose, it is also useful to formulate the axiom in two
parts: \medskip

Consider two subsets $B,$ $B^{\prime }\in \mathcal{B}\ $such that $B\subset
B^{\prime }.$

(a)\ Let $x,y\in B,\ x\neq y$ with \thinspace $x\in C\left( B\right) $ then
\thinspace $y\notin C\left( B^{\prime }\right) .\ $

(b) Let $C\left( B^{\prime }\right) \cap B\neq \varnothing $ then $C\left(
B\right) =C\left( B^{\prime }\right) .\medskip $

The intuition for (a) is that as we enlarge the set of items from $B$ to $%
B^{\prime }$ a rational decision-maker never chooses from $B^{\prime }$ an
item that is available in $B$ but is not chosen.\ The intuition for (b) is
that as we reduce the set of alternatives an item chosen in the large set,
is also chosen in the smaller set containing that item and no item
previously not chosen becomes chosen.\footnote{%
Conditions (a) and (b) are equivalent to C2 and C4 in Arrow (1959).}

It can be shown that in the classical context $(b)$ implies $(a)\ $(see
Arrow (1959))$.$ In our context where choice experiments can be
non-commuting, we do not have such an implication$.$ Moreover, the notion of
choice function is not appropriate because it implicitly assumes the
commutativity of choice\textbf{\ }(see below).\textbf{\ }Since our purpose
is to investigate this issue explicitly, we express the axiom more
immediately in terms of (observable) choice behavior in the following way
and we call it WARP': \medskip

Consider two subsets$\ B,$ $B^{\prime }\in \mathcal{B}$\ with $B\subset
B^{\prime }$.

1) Suppose the agent chooses from $B$\ some element $x$. In a next
subsequent measurement of $B^{\prime }$\ the outcome of the choice is not in 
$B\setminus \{x\}$.

2) Suppose the agent chooses from $B^{\prime }$\ some element $x$ that
belongs to $B$. In a next subsequent measurement of $B$\ the agent's choice
is $x$.\medskip

Points (1) and (2) capture the classical intuitions about rational choice
behavior associated with (a) and (b) above. A distinction with the standard
formulations of WARP is that we do not refer to a choice function but to
choice behavior and that our axiom only applies to two subsequent choices.
We shall see below that in the classical context where choices commute WARP'
is equivalent to WARP. That is not longer true if we allow for some choices
not to commute.

\ 

We now investigate the choice behavior of a type indeterminate agent under
assumptions A1 and A2 above i.e., under the assumption that choices from
small subsets satisfy the property of first-kindness and that they respect
WARP'. We consider in turn two cases\textbf{.} In the first one, we assume
that all the \textit{DS} considered pairwise\textbf{\ }commute. In the
second one, we allow for non-commuting \textit{DS}. In the following we
define \textquotedblleft small\textquotedblright\ subsets as subsets of size
3 or less.

\ 

\textit{Case 1}

The assumption here is that all experiments of choice from small subsets are
compatible with each other, i.e., the corresponding observables commute. A
first implication of the commutativity of choice experiments is that for any 
$B\subseteq X,$\ the outcome of the choice experiment $B$\ only depends on
\thinspace $B$ (whatever was done before the choice in $B$ will always be
the same and in particular the outcome does not depend on choices
experiments that were performed before)$.$\ Therefore, we can express the
outcome in terms of \textit{a choice function} $C\left( X\right)
:B\rightarrow x$ that associates to each $B\in \mathcal{B}\ $\ a chosen item 
$x\in B.\ $A second consequence of commutativity is that the restriction of
WARP' to two subsequent choices is inconsequential. This is because the
outcome of any given choice experiment must be the same in any series of two
consecutive experiments. In particular, WARP' implies the transitivity of
choices. This can easily be seen we taking an example on a subset of 3 items
and performing choice experiments on pairs in different orders.\footnote{%
Consider the choice set $\left\{ a,b,c\right\} .$\ We do the following two
series of experiments. In the first series we let the agent choose from $%
\left\{ a,b\right\} $, then\ from $\left\{ b,c\right\} \ $and last from $%
\left\{ a,b,c\right\} .$In the second series the agent first choose from $%
\left\{ b,c\right\} $ then from $\left\{ a,b\right\} ~$and$\ $last from $%
\left\{ a,b,c\right\} $.\ Assume the outcomes of the two first experiments
are $a$ and $\ b\ $then\ by WRAP' the third choice maybe either $a,\ $or $%
b.\ $In the second series, the outcomes must be (because of commutativity) $%
\ b$ and $a\ $respectively.$\ $So the third choice may be either $a$ or $c.$%
\ By commutativity the outcome of $\left\{ a,b,c\right\} $ must be the same
in the two series which uniquelly selects $a$ so a choice behavior
respecting WRAP' is transitive when the choices commute.}\textbf{.} So for
the case choice experiments commute, WARP' is equivalent to WARP and we
recover the standard results. We know that if $\left( \mathcal{B},\mathcal{\ 
}C\left( .\right) \right) \ $is\ a choice structure satisfying WARP' and
defined for $\mathcal{B\ }$including all subsets of $X\ $of up to three
elements,\ then there exists a rational preference relation that
rationalizes choice behavior.\footnote{%
See e.g., Mas-Colell\ et al. (1995) p.13.} It is therefore natural to
identify the states (types) of the individual with the preference orders
that rank all the elements of $X.\ $The type space $\mathcal{H}$ has
dimension \thinspace $\left\vert X\right\vert !.$ We obtain the well-known
classical model.

\ 

In this model all \textit{DS }are coarse measurements of the type (the
preference order) i.e., their outcomes are degenerated eigenvalues (see
Section 2.3.1)$.\ $When all$\ $\textit{DS}$\ $commute and satisfy WARP', the
type indeterminate rational agent behaves in all respects as a classical
rational agent. This should not surprise us since we know that when
observables commute they can be merged into a single observable and the
TI-model is equivalent to the classical model.

\smallskip\ 

\textit{Case 2}

We now consider a case where some \textit{DS} do not commute. We need to
emphasize that, in contrast with the commuting case, a large amount of
non-commuting models are possible. The models differ from each other
according to which \textit{DS} commute and which do not. We investigate here
a simple example that illustrates interesting issues.

Assume the set $X$ consists of four items: $a,b,c$ and $d$. As before, any
subset $A\subset X$ consisting of 3 elements or less is associated with a 
\textit{DS} and we assume that any two consecutive choices made from small
subsets respect WARP'. Our non-commuting model is defined by the following
assumption:\medskip

\textit{(nc)} Choices out of $A$ and $B$ commute \textit{if and only if} $%
A\cup B\neq X.$ \medskip

In particular \textit{DS} associated with different triples e.g., a choice
out of $\left\{ a,b,c\right\} $ and a choice out of $\left\{ b,c,d\right\} \ 
$are represented by observables that \textit{do not} commute with each
other. So for instance the agent may choose $a$ out of $\left\{
a,b,c\right\} ,$ $b$ out of $\left\{ b,c,d\right\} $\ and thereafter $c$\
out of $\left\{ a,c\right\} .$ This may happened because the order
(partially) revealed by the choice in $\left\{ a,b,c\right\} $ has been
modified by the performance of the $\left\{ b,c,d\right\} $ choice
experiment. $\ $This choice behavior satisfies WARP' but violates
transitivity$.\ $Nevertheless, the satisfaction of WARP' induces a certain
amount of consistency in behavior. In particular, any observable $A\;$with$\
A\subset B$ where $B$ is a triple, commutes with$\ B\ ($since$\ A\cup
B\subset X)$. We may perform choice experiments in pairs and in triples so
as to elicit a preference relation on each triple \textit{taken separately}.
It is therefore natural to identify the types with preferences orders on
triples. Since observables representing choice experiments on different
triples do not commute with each other, they are \textit{alternative} ways
to measure the individual's preferences. The type space representing the
individual is a six dimensional Hilbert space because there are six ways to
rank three items. There are four alternative bases spanned by the
eigenvectors corresponding to the six possible rankings in each one of the
four triples.

\ 

In this model the agent only has preferences over triples of items. One may
wonder what happens when she must make a choice out of the set $\left\{
a,b,c,d\right\} .$ Many scenarios are possible. Suppose that the individual
proceeds by making a series of two measurements. For instance, she first
selects from a pair followed by a choice from a triple consisting of the
first selected item and the two remaining ones. We can call such a behavior
"procedural rational" because she acts \textit{as if} she had preferences
over the four items.

As an example consider the following line of events and for the ease of
presentation we let the \textit{DS }associated with the set $\left\{
a,b,c\right\} $ be denoted $abc$ and similarly for the other $\mathit{DS}.$
Suppose the individual just made a choice in $\left\{ a,b,c\right\} $\ and
picked up $c.\ $So her initial state is some superposition of type $[c>b>a]$
and type $[c>a>b].$ We now ask her to choose from $\left\{ a,b,c,d\right\} .$%
\ Assume the individual first plays $ab$ then $bcd$\ (which by assumption of
procedural rationality means that the outcome of $ab\ $is $b)$.\footnote{%
We remind that "playing a \textit{DS"} means performing the corresponding
measurement (see section 2.2).}$\ $The choice of $b$ from $\left\{
a,b\right\} \ $changed the type of the individual. In particular, suppose
the\ $ab$ observable is considered as a coarse measurement of $abd\mathit{.\ 
}$The new type, expressed in the basis of preference orders on $\left\{
a,b,d\right\} \mathit{,}$ is a superposition of $\ $types$\ \left[ b>a>d%
\right] ,\ \left[ b>d>a\right] \ $and $\left[ d>b>a\right] .\ $In order to
make her choice out of $\left\{ a,b,c,d\right\} ,$ she now plays$\ bcd.$ The
result of that last measurement (performed on the individual of the type
resulting from the first measurement) is therefore $b$ with positive
probability. But this violates the principle of independence of irrelevant
alternative (IIA)$.\ $She effectively selects $b$ from $\left\{
a,b,c,d\right\} \ $while she initially picked up $c$ in $\left\{
a,b,c\right\} $ where $b$ was available.\textbf{\ }Yet, it is easy to check
that in this example any two consecutive choices satisfy the WARP' so the
choice behavior of our type indeterminate agent is "rational". \ 

\ \smallskip

We can now draw some first conclusions from this short exploration. If we
demand that the agent's choice behavior respects the Weak Axiom of Revealed
Preference (in our WARP' formulation) and conforms to procedural
rationality, then

i. when all the \textit{DS} associated with subsets of the universal set
commute, a type indeterminate agent does not distinguish herself from a
classical agent;

ii. when some \textit{DS} associated with subsets of the universal set do
not commute with each other, a type indeterminate agent does \textit{not}
have a preference order over the universal set $X.$\ But she may have
well-behaved preferences over subsets of $X.\ $Generally, she does not
behave as a classical rational agent.\textit{\ }In particular, her behavior
may exhibit standard phenomena of preference reversal.

\ 

In this example the distinction between the classical rational and the type
indeterminate rational agent is only due to the non-commutativity of \textit{%
DS\ }associated with different subsets of items. This distinctive feature of
the TI-model of choice delivers a formulation of bounded rationality in
terms of the impossibility to compare and order all items \textit{%
simultaneously}. Non-commutativity also implies that as the agent makes a
choice her type (preferences) changes. A type indeterminate agent does not
have a fixed type (preferences). Instead her preferences are shaped in the
process of elicitation as proposed by Kahneman and Tversky (2000).

\subsection{Examples}

In this section we demonstrate how type indeterminacy can explain two
well-documented examples of so-called behavioral anomalies. With these
examples we want to suggest that one contribution of our approach is to
provide a unified framework which can accommodate a variety of behavioral
anomalies. These anomalies are currently explained by widely different
theories.

\subsubsection{Cognitive dissonance}

The kind of phenomena we have in mind can be illustrated as follows\textbf{.}
Numerous studies show that employees in risky industry (like nuclear plants)
often neglect safety regulations. The puzzle is that before moving into the
risky industry those employees were typically able to estimate the\textbf{\ }%
risks correctly. They were reasonably averse to risk and otherwise behaved
in an ordinary rational manner.

Psychologists developed a theory called cognitive dissonance (CD) according
to which people modify their beliefs or preferences in response to the
discomfort arising from conflicting beliefs or preferences. In the example
above, they identify a dissonance as follows. On the one hand, the employee
holds an image of himself as \textquotedblright a smart
person\textquotedblright\ and on the other hand he understands that he
deliberately chose to endanger his health (by moving to the risky job),
which is presumably \textquotedblright not smart\textquotedblright . So in
order to cancel the dissonance, the employee decides that there is no danger
and therefore no need to follow the strict safety regulation.

We propose a formal model that is very much in line with psychologists'
theory of cognitive dissonance. We shall compare two scenarios involving a
sequence of two non-commuting Decision Situations each with two options.
Let\ $A$ be a \textit{DS }about jobs with options$\;a_{1}$ and $a_{2}$\
corresponding to taking a job with a dangerous task (adventurous type) and%
\textbf{\ }respectively\ staying with the safe routine (habit-prone type).\
Let $B\;$be a \textit{DS} about the\textbf{\ }willingness to use safety
equipement with$\;$choices$\ b_{1}\ $(risk-averse type) and$\ b_{2}\ $%
(risk-loving type) corresponding to\ the choice of using and\textbf{\ }%
respectively not using the safety equipment.

\medskip

\textit{First scenario}:\ The dangerous task is introduced in an existing
context. It is \textit{imposed} on the workers. They are \textit{only} given
the choice to use or not to use the safety equipment ($B)$. We write the
initial state of the worker in terms of the eigenvectors of observable $A$:

\begin{equation*}
\left\vert \psi \right\rangle =\lambda _{1}\left\vert a_{1}\right\rangle
+\lambda _{2}\left\vert a_{2}\right\rangle ,\ \lambda _{1}^{2}+\lambda
_{2}^{2}=1.
\end{equation*}%
We develop the eigenvectors of \textit{A} on the eigenvectors of \textit{B: }
\begin{eqnarray*}
\left\vert a_{1}\right\rangle &=&\mu _{11}\left\vert b_{1}\right\rangle +\mu
_{21}\left\vert b_{2}\right\rangle , \\
\left\vert a_{2}\right\rangle &=&\mu _{21}\left\vert b_{1}\right\rangle +\mu
_{22}\left\vert b_{2}\right\rangle .
\end{eqnarray*}%
We now write the state in the basis of the $B$ operator: 
\begin{equation*}
\left\vert \psi \right\rangle =\left[ \lambda _{1}\mu _{11}+\lambda _{2}\mu
_{21}\right] \left\vert b_{1}\right\rangle +\left[ \lambda _{1}\mu
_{12}+\lambda _{2}\mu _{22}\right] \left\vert b_{2}\right\rangle .
\end{equation*}%
The probability that a worker chooses to use the safety equipment is 
\begin{eqnarray}
p_{B}\left( b_{1}\right) &=&\ \left\langle b_{1}\right\vert \left. \psi
\right\rangle ^{2}=\left[ \lambda _{1}\mu _{11}+\lambda _{2}\mu _{21}\right]
^{2}  \label{probb1} \\
&=&\lambda _{1}^{2}\mu _{11}^{2}+\lambda _{2}^{2}\mu _{21}^{2}+2\lambda
_{1}\lambda _{2}\mu _{11}\mu _{21}.  \notag
\end{eqnarray}

\textit{Second scenario:} First $A$ then $B$. The workers choose between
taking a new job with a dangerous task or staying with the current safe
routine. Those who chose the new job then face the choice between using
safety equipment or not. Those who turn down the new job offer are asked to
answer a questionnaire about their willingness to use the safety equipment
for the case they would be working in the risky industry. The ex-ante
probability for observing $b_{1}$ is 
\begin{eqnarray}
p_{BA}\left( b_{1}\right) &=&p_{A}\left( a_{1}\right) p_{B}\left( \left.
b_{1}\right\vert a_{1}\right) +p_{A}\left( a_{2}\right) p_{B}\left( \left.
b_{1}\right\vert a_{2}\right)  \label{probb2} \\
&=&\lambda _{1}^{2}\mu _{11}^{2}+\lambda _{2}^{2}\mu _{21}^{2}.  \notag
\end{eqnarray}%
The phenomenon of cognitive dissonance can now be formulated as the
following inequality 
\begin{equation*}
p_{BA}\left( b_{1}\right) <p_{B}\left( b_{1}\right) ,
\end{equation*}%
which occurs in our model when $2\lambda _{1}\lambda _{2}\mu _{11}\mu
_{21}>0.$\footnote{%
We note that $p_{BA}\left( b_{1}\right) $ includes the probability of a
choice of safety measures \textit{both} in the group that chose the risky
job and in the group that chose the safe job. This guarantees that we
properly distinguish between the CD effect (change in attitude) and the
selection bias.}$\;$We next show that interference effects may be
quantitatively significant.

\textit{Numerical} \textit{example}

Assume for simplicity that $\left\vert \psi \right\rangle =\left\vert
b_{1}\right\rangle $ which means that everybody in the first scenario is
willing to use the proposed safety equipment. Let$\;prob\left(
a_{1}\left\vert \psi \right. \right) =0.25,\;$and$\;prob\left(
a_{2}\left\vert \psi \right. \right) =.75$ so $\left\vert \lambda
_{1}\right\vert =\sqrt{0.25},\;$and$\;\left\vert \lambda _{2}\right\vert =%
\sqrt{0.75}.$\ It is possible to show that in this case we have $\left\vert
\mu _{11}\right\vert =\sqrt{0.25},\;$and$\;\left\vert \mu _{21}\right\vert =%
\sqrt{0.75}.$\footnote{%
This uses the fact that $\left( 
\begin{tabular}{ll}
$\mu _{11}$ & $\mu _{22}$ \\ 
$\mu _{21}$ & $\mu _{22}$%
\end{tabular}%
\right) $ is a rotation matrix.}$\ $We now compute $p_{B}^{{}}(b_{1})$ $\ $%
using the formula in (\ref{probb1}) and recalling that $\left\vert \psi
\right\rangle =\left\vert b_{1}\right\rangle \;$(so$\ p_{B}^{{}}(b_{1})=1):$ 
\begin{eqnarray}
1 &=&\left\langle b_{1}\right. \left\vert \psi _{{}}\right\rangle
^{2}=\lambda _{1}^{2}\mu _{11}^{2}+\lambda _{2}^{2}\mu _{21}^{2}+2\lambda
_{1}\lambda _{2}\mu _{11}\mu _{21}  \label{sce1} \\
&=&0.0625+0.5625+2\lambda _{1}\lambda _{2}\mu _{11}\mu _{21}=0.625+2\lambda
_{1}\lambda _{2}\mu _{11}\mu _{21}.  \notag
\end{eqnarray}%
which\ implies that\ the interference effect is positive and equal to $%
1-0.625=0.375$. We note that it amounts to a third of the probability.

Under the second scenario the probability for using safety equipment is
given by the formula in (\ref{probb2}) i.e., it is the same sum as in (\ref%
{sce1}) without the interference term: 
\begin{equation*}
p_{BA}\left( b_{1}\right) =0.625.
\end{equation*}%
So we see that our TI-model "delivers" cognitive dissonance: $p_{BA}\left(
b_{1}\right) <p_{B}\left( b_{1}\right) .$

The key assumption that drives our result is that the choice between jobs
and the choice between using or not using the safety equipment are
measurements of two incompatible type characteristics (represented by two
non-commuting observables). A possible behavioral interpretation is as
follows. The job decision appeals to an abstract perception of risk, while
the decision to use the safety equipment appeals to an emotional perception
of concrete risks. In this interpretation, our assumption is that the two
modes of perceptions are incompatible. This is consistent with evidence that
shows a gap between perceptions of one and the same issue when the agent is
in a "cold" (abstract) state of mind compared to when she is in a "hot"
emotional state of mind (see for instance Loewenstein 2005).\ 

\textit{Comments}

In their article from 1981 Akerlof and Dickens explain the behavior
attributed to cognitive dissonance in terms of a rational choice of beliefs.
Highly sophisticated agents choose their beliefs to fit their preferences.%
\footnote{%
Akerlof and Dickens allow workers to freely choose beliefs (about risk) so
as to optimize utility which includes psychological comfort. The workers are
highly rational in the sense that when selecting beliefs, they internalize
their effect their own subsequent bounded rational behavior.} They are fully
aware of the way their subjective perception of the world is biased and yet
they keep to the wrong views. This approach does explain observed behavior
but raises serious questions as to what rationality means. The type
indeterminacy approach is consistent with psychologists' thesis that
cognitive dissonance prompts a change in preferences (or attitude). We view
its contribution as follows. First, the TI-model provides a model that
explains the appearance of cognitive dissonance. Indeed, if coherence is
such a basic need, as proposed by L. Festinger and his followers, why does
dissonance arise in the first place? In the TI-model dissonance arises when
resolving indeterminacy in the first \textit{DS} because of the
`limitations' on possible psychological types (see Section 2.3.2). Second,
the TI-model features a \textit{dynamic process} such that the propensity to
use safety measures is actually altered (reduced) as a consequence of the 
\textit{act of choice}.\ This dynamic effect is reminiscent of
psychologists' \textquotedblleft drive-like property of
dissonance\textquotedblright\ which leads to a change in attitude.\ 

\subsubsection{Framing Effects}

When alternative descriptions of one and the same decision problem induce
different decisions from agents we talk about "framing effects". Below we
attend a well-known experiment which showed that two alternative
formulations of the Prisoner Dilemma (the standard presentation in a 2 by 2
matrix and a presentation in a decomposed form, see below) induced
dramatically different rates of cooperation.

Kahneman and Tversky (2002, p. xiv) address framing effects using a
two-steps (nonformal) model of the decision-making process. The first step
corresponds to the construction of a representation of the decision
situation. The second step corresponds to the evaluation of the choice
alternatives. The crucial point is that\textit{\ \textquotedblleft the true
objects of evaluation are neither objects in the real world nor verbal
description of those objects; they are mental
representations.\textquotedblright\ }To capture this feature we suggest
modelling the "process of constructing a representation" in a way similar to
the process of constructing preferences, i.e., as a measurement performed on
the state of the agent. This is consistent with psychology that treats
attitudes, values (preferences), beliefs, and representations as mental
objects of the same kind.\ 

In line with Kahneman and Tversky we describe the process of decision-making
as a sequence of two steps. The first step consists of a measurement of the
agent's mental representation of the decision situation.\textit{\ }The
second step corresponds to a measurement of the agent's preferences. Its
outcome is a choice. Note that here\textbf{\ }we depart from standard
decision theory. We propose that agents do not always have a \textit{unique}
representation of a decision situation, which can be revealed when thinking
about it. Instead, uncertainty about what the decision situation actually is
about can be resolved in a variety of ways some of which may be reciprocally
incompatible. The decision situation itself is, as in standard theory,
defined by the monetary payoffs associated with the choices i.e., in a
unique way. The \textit{utility} associated with the options generally
depends both on the representation and the monetary payoffs.

To illustrate this approach we revisit the experiment reported in Pruitt
(1970) and discussed in Selten (1998). Two groups of agents are invited to
play a Prisoners' Dilemma. The game is presented to the first group in the
usual matrix form with the options labelled $1_{G}$\ and$\;2_{G}$ (instead
of $C$ and $D,$ presumably to avoid associations with the suggestive terms
`cooperate' and `defect'):

\bigskip \medskip \medskip $%
\begin{tabular}{|l|l|l|}
\hline
& $\left[ C\right] $ & $\left[ D\right] $ \\ \hline
$1_{G}$ & $%
\begin{array}{cc}
& 3 \\ 
3 & 
\end{array}%
$ & $%
\begin{array}{cc}
& 4 \\ 
0 & 
\end{array}%
$ \\ \hline
$2_{G}$ & $%
\begin{array}{cc}
& 0 \\ 
4 & 
\end{array}%
$ & $%
\begin{array}{cc}
& 1 \\ 
1 & 
\end{array}%
$ \\ \hline
\end{tabular}%
\ \ $

\bigskip The game is presented to the second group in a decomposed form as
follows:

\ 

\ $\ 
\begin{tabular}{|l|l|l|}
\hline
& For me & For him \\ \hline
1$_{G}$ & $0$ & 3 \\ \hline
2$_{G}$ & 1 & 0 \\ \hline
\end{tabular}
\ $

\bigskip

The payoffs are computed as the sum of what you take for yourself and what
you get from the other player. So for instance if player 1 plays 1$_{G}\ $%
and player 2 plays 2$_{G}$,\ player 1 receives $0$ from his own play and $0$
from the other's play) which sums to 0. Player 2 receives 1 from his own
play and 3 from player 1's play which sums to 4. So we recover the payoffs
(0,4) associated with the play of Cooperate for player 1 and Defect for
player 2. Game theoretically it should make no difference whether the game
is presented in a matrix form or in a decomposed form. Pruitt's main
experimental result is that one observes dramatically more cooperation in
the game presented in decomposed form.

\medskip

We now provide a possible TI-model for this situation. Let us consider a
two-dimensional state space and a sequence of two measurements. The first
measurement determines the mental representation of the \textit{DS.}$\ $We
call $A$ the (representation) observable corresponding to the matrix
presentation. It has two non-degenerated eigenvalues denoted $\left\vert
a_{1}\right\rangle $ and $\left\vert a_{2}\right\rangle .$\ Similarly, $B$
is the observable corresponding to the decomposed presentation with two
eigenvalues $\left\vert b_{1}\right\rangle $ and $\left\vert
b_{2}\right\rangle $. If $\left\vert \psi \right\rangle $ is the initial
state of the agent, we can write $\left\vert \psi \right\rangle \ =\alpha
_{1}\left\vert a_{1}\right\rangle +\alpha _{2}\left\vert a_{2}\right\rangle $
or $\left\vert \psi \right\rangle \ =\beta _{1}\left\vert b_{1}\right\rangle
+\beta _{2}\left\vert b_{2}\right\rangle .\ $The second measurement i.e.,
the decision observable is unique, we called it$\ G$. $G$ has also two
eigenvalues denoted $1_{G}$ and $2_{G}.\;$

For the sake of concreteness we may think of the four alternative mental
representations as follows\footnote{%
This is only meant as a suggestive illustration.}:

$\left\vert a_{1}\right\rangle :\;G$ is perceived as an (artificial)
small-stake game;

$\left\vert a_{2}\right\rangle :\;G$ is perceived by analogy as a real life
PD-like situation (often occurring in a repeated setting).

$\left\vert b_{1}\right\rangle :\;G$ is perceived as a test of generosity;

$\left\vert b_{2}\right\rangle :\;G$ is perceived as a test of smartness.

$\;$

When confronted with a presentation of the $\mathit{DS}$ the agent forms her
mental representation of the \textit{DS} which prompts a change in her state
from $\left\vert \psi \right\rangle \ $to some $\left\vert
a_{i}\right\rangle $ (if the frame is $A)\ $or $\left\vert
b_{j}\right\rangle \;i,j=1,2\ ($if the frame is $B)$. The new state can be
expressed in terms of the eigenvectors of the decision situation: $%
\left\vert a_{i}\right\rangle =\gamma _{1i}\left\vert 1_{G}\right\rangle
+\gamma _{2i}\left\vert 2_{G}\right\rangle $ or $\ \left\vert
b_{j}\right\rangle =\delta _{1j}\left\vert 1_{G}\right\rangle +\delta
_{2j}\left\vert 2_{G}\right\rangle .\;$

We can now formulate the framing effect as the following difference 
\begin{equation}
p_{GA}\left( i_{G}\right) \neq p_{GB}\left( i_{G}\right) ,\;\text{ }i=1,2.%
\text{ }  \label{framing}
\end{equation}%
Using our result in Section 2.3.2 we get 
\begin{equation*}
p_{GA}\left( 1_{G}\right) =p_{G}\left( 1_{G}\right) -2\alpha _{1}\gamma
_{11}\alpha _{2}^{{}}\gamma _{12}^{{}}\text{ and\ }p_{GB}\left( 1_{G}\right)
=p_{G}\left( 1_{G}\right) -2\beta _{1}\delta _{11}\beta _{2}^{{}}\delta
_{12}^{{}}
\end{equation*}%
where $p_{G}\left( 1_{G}\right) $ is the probability of choosing 1$\ $in\ an
(hypothetical) unframed situation.\ So we have a framing effect whenever $%
2\alpha _{1}\gamma _{11}\alpha _{2}^{{}}\gamma _{12}^{{}}\neq 2\beta
_{1}\delta _{11}\beta _{2}^{{}}\delta _{12}^{{}}.\ $

The experimental results discussed in Selten (1998)\ namely that the
decomposed presentation induces more cooperation require than $2\alpha
_{1}\gamma _{11}\alpha _{2}^{{}}\gamma _{12}^{{}}-2\beta _{1}\delta
_{11}\beta _{2}^{{}}\delta _{12}^{{}}>0$. The inequality says that the
interference term for $1_{G}$ is larger in the standard matrix presentation $%
A$ than in the decomposed form corresponding to the $B$ presentation.\ In
order to better understand the meaning of this difference we shall consider
a simple numerical example.

Set $\alpha _{1}=\sqrt{.8},\ \alpha _{2}=\sqrt{.2,\ }\beta _{1}=\sqrt{0.4},\
\beta _{2}=\sqrt{0.6}.$\ The key variables are the correlations between the
"representation types" i.e., $\left\vert a_{i}\right\rangle $ or $\left\vert
b_{j}\right\rangle $ and the "game type" $1_{G},$ i.e., the numbers $\gamma
_{11},\gamma _{12}^{{}}$ and $\delta _{11},\ \delta _{12}^{{}}.$ We propose
that $\gamma _{11}=\sqrt{.3}$ which is interpreted as when the agent views
the game as a small stake game she plays $1_{G}$ with probability .3.\
Similarly $\gamma _{12}=\sqrt{.7}$ which means that when the agent perceives
the game by analogy with real life, she "cooperates" with probability .7. In
the alternative presentation $B,$ we propose that $\delta _{11}=1,$ i.e.,
when the game is perceived as a test of generosity, the agent cooperates
with probability 1. When the game is perceived as a test of smartness $%
\delta _{12}=0$ (because the agent views the play of $1_{G}$\ as plain
stupid)$.$ Computing these figures, we get

\begin{equation*}
2\alpha _{1}^{{}}\gamma _{11}\alpha _{2}^{{}}\gamma _{11}-2\beta _{1}\delta
_{11}\beta _{2}^{{}}\delta _{12}^{{}}=0.366-0>0
\end{equation*}

In the $A$ presentation the contribution of both the $\left\vert
a_{1}\right\rangle $ and the $\left\vert a_{2}\right\rangle \ $type is
positive and significant. When the agent is indeterminate both types
positively contribute reinforcing each other. In contrast in the $B$
presentation the contribution from $\left\vert b_{2}\right\rangle $ is null
so there is no interference between the types. When the agent determines
herself i.e., selects a representation, the contribution from indeterminacy
is lost and that loss is positive with $A$ while it is null with $B$.
Therefore, the probability for $1_{G}\ $when the game is presented in the
matrix form is larger (here by .36) than in the game presented in the
decomposed form.

\ \ 

\textit{Comments}

Selten proposes a `bounded rationality' explanation: players make a
superficial analysis and do not perceive the identity of the games presented
under the two forms. Our approach is closer to Kahneman and Tversky who
suggest that prior to the choice, a representation of the decision situation
must be constructed. The TI-model provides a framework for `constructing' a
representation such that it delivers framing effects in choice behavior. Of
course framing effects can easily be obtained when assuming that the mental
images are incomplete or biased. In the TI-model we do not need to appeal to
such self-explanatory arguments. In the TI-model framing effects arise as a
consequence of (initial) indeterminacy of the agent's representation of the
decision situation. Since alternative (non-commuting) presentations are
equally valid and their corresponding representations (eigenvalues) equally
informative, highly rational agent can exhibit framing effects.

\section{Discussion}

In this section we briefly discuss some formal features of our model. 
\footnote{%
For a systematic investigation of the mathematical foundations of the HSM in
view of their relevance for social sciences see Danilov and
Lambert-Mogiliansky (2005).}

Our\textbf{\ }approach to decision-making yields that the type of the agents
rather than being exogenously given, emerges as the outcome of the
interaction between the agent and the decision situations. This is modelled
by letting a decision situation be represented by an operator (observable).
Decision-making is modelled as a measurement process. It projects the initial%
\textbf{\ }state of the agent\textbf{\ }into the subspace of the state
(type) space associated with the eigenvalue corresponding to the choice
made. Observables may either pairwise commute or not. When the observables
commute, the corresponding type space has the properties of the Harsanyi
type space. From a formal point of view this reflects the fact that all
(pure) types are then mutually exclusive.\footnote{%
We say that a type is pure when it is obtained as the result of a complete
measurement i.e., the measurement of a complete set of commuting observables
(CSCO).}\ When the observables do not commute, the associated pure types are
not all mutually exclusive. Instead, an agent who is in a pure state after
the measurement of an observable will be in a different pure state after the
measurement of another observable that is incompatible with the first one.
As a consequence, the type space cannot be associated with a classical
probability space and we obtain an irreducible uncertainty in behavior. The
Type Indeterminacy model provides a framework where we can deal both with
commuting and \textbf{\ }non-commuting observables. In the TI-model any type
(state ) corresponds to a probability measure on the type space which allows
to make predictions about the agent's behavior. It is in this sense that the
TI-model generalizes Harsanyi's approach to uncertainty.

The more controversial feature of the TI-model\ as a framework for
describing human behavior is related to the modelling of the impact of
measurement on the state i.e., how the type of the agent changes with
decision-making. The rules of change are captured in the geometry of the
type space and in the projection postulate. It is more than justified to
question whether this seemingly very specific process should have any
relevance to Social Sciences.

It has been shown that the crucial property that gives all its structure to
the process of change can be stated as a "minimal perturbation principle".
The substantial content of that principle is that we require that when a
coarse \textit{DS} resolves some uncertainty about the type of an agent, the
remaining uncertainty is left unaffected. Recall our example in section 3.1
case 2. When the agent chooses an item out of a subset $A$\ of 3 items, this
prompts a resolution of some uncertainty. The type is projected into the
eigenspace spanned by the two orderings consistent with the choice made. The
minimal perturbation principle says that uncertainty relative to the
ordering of the two remaining items is the same as before. In behavioral
terms this can be expressed as follows. When confronted with the necessity
to make a choice, the agent only "makes the effort" to select her preferred
item while leaving the order relationship between the other items uncertain
as initially.

It may be argued that the minimal perturbation principle is quite demanding.
Returning again to our example, if the mental processes involved in the
search for the preferred items fully upset the initial state, the principle
is violated. It could also be argued that the mental processes involved in
decision-making determine the whole ranking. That would also violate the
minimal perturbation principle. This short discussion suggests that
selecting a good TI-model requires careful thinking and possibly some trial
and error.

We do not expect the Type Indeterminacy model to be a fully realistic
description of the human behavior rather we propose it as an idealized model
of agents characterized by the fact that their type change with
decision-making. In particular some features of the TI-model, like the
symmetry of the correlation matrix in simple examples may seem very
constraining from a behavioral point of view. Consider two \textit{DS} with
two options e.g., the Prisoner Dilemma and the Ultimatum game (with option
"share fairly" and "share egoistically"). Assuming those \textit{DS} have
nondegenerated eigenvalues, the symmetry of correlation matrix means that
the probability to play e.g. defect after having played say egoist is
exactly the same as the probability to play egoist after having played
defect. We do not expect this kind of equality to hold in general.\footnote{%
Note that a failure of this inequality to hold may be due to the fact that
the \textit{DS }have some degenerated eigenvalues. In fact, this is likely
to be the case in Social Sciences since we often deal with rather coarse
measurements of the agent's type.} Nevertheless, keeping in mind some
reservations as to its realism, our view is that the Type Indeterminacy
model can provide a fruitful framework for analyzing, explaining and
predicting human behavior. Clearly much additional work is needed to extend
of the TI-model to strategic and repeated decision-making. We are currently
exploring this second stage of our research program.

\ 

As a final remark it should be emphasized that not all instances of
non-commutativity in choice behavior call for Hilbert space modelling.
Theories of addiction feature effects of past choices on future preferences.
And in standard consumer theory, choices do have implications for future
behavior, i.e., when goods are substitutes or complements. But in those
cases we \textit{do} expect future preferences to be affected by the
choices. The Type Indeterminacy model of decision-making can be useful when
we expect choice behavior to be consistent with the standard probabilistic
model, because nothing justifies a modification of preferences. Yet, actual
behavior contradicts those expectations.

\section{Appendix: Elements of Quantum Mechanics}

\subsection{States and Observables}

In Quantum Mechanics the state of a system is represented by a vector $%
\left\vert \psi \right\rangle $ in a Hilbert space $\mathcal{H}$. According
to the superposition principle, every complex linear combination of state
vectors is a state vector. A Hermitian operator called an observable is
associated to each physical property of the system.\ 

\bigskip

\emph{Theorem 1}

A Hermitian operator$\;A$ has the following properties:

- Its eigenvalues are real.

- Two eigenvectors corresponding to different eigenvalues are orthogonal.

- There is an orthonormal basis of the relevant Hilbert space formed with
the eigenvectors of $A.$

\bigskip

Let us call $\left| v_{1}\right\rangle ,\left| v_{2}\right\rangle
,...,\left| v_{n}\right\rangle $ the normalized eigenvectors of $A$ forming
a basis of $\mathcal{H}$. They are\ associated with eigenvalues $\alpha
_{1,}\alpha _{2},...,\alpha _{n}$, so: $A\left| v_{i}\right\rangle =\alpha
_{i}\left| v_{i}\right\rangle $. The eigenvalues can possibly be
degenerated, i.e., for some $i$ and $j$, $\alpha _{i}=\alpha _{j\text{.}}$
This means that there is more than one linearly independent eigenvector
associated with the same eigenvalue. The number of these eigenvectors
defines the degree of degeneracy of the eigenvalue which in turn defines the
dimension of the eigensubspace spanned by these eigenvectors. In this case,
the orthonormal basis of $\mathcal{H}$ is not unique because it is possible
to replace the eigenvectors associated to the same eigenvalue by any complex
linear combination of them to get another orthonormal basis. When an
observable $A$ has no degenerated eigenvalue there is a unique orthonormal
basis of $\mathcal{H}$ formed with its eigenvectors. In this case (see
below), it is by itself a Complete Set of Commuting Observables.

\bigskip

\emph{Theorem 2}

If two observables $A$ and $B$ commute there is an orthonormal basis of $%
\mathcal{H}$ formed by eigenvectors common to $A$ and $B$.

\bigskip

Let $A$ be an observable with at least one degenerated eigenvalue and $B$
another observable commuting with $A.$ There is no unique orthonormal basis
formed by $A$ eigenvectors. But there is an orthonormal basis of the
relevant Hilbert space formed by eigenvectors common to $A$ and $B$. By
definition, $\left\{ A,B\right\} $ is a Complete Set of Commuting
Observables (CSCO) if this basis is unique. Generally, a set of observables $%
\left\{ A,B,...\right\} $ is said to be a CSCO if there is a unique
orthonormal basis formed by eigenvectors common to all the observables of
the set.

\bigskip

\subsection{Measurements}

An observable $A\ $is associated to each physical property of a system $S$.
Let $\left| v_{1}\right\rangle ,\left| v_{2}\right\rangle ,...,\left|
v_{n}\right\rangle $ be the normalized eigenvectors of $A$ associated
respectively with eigenvalues $\alpha _{1,}\alpha _{2},...,\alpha _{n}$ and
forming a basis of the relevant state space. Assume the system is in the
normalized state $\left| \psi \right\rangle $. A\ measurement of $A$ on $S$
obeys the following rules, collectively called `Wave Packet Reduction
Principle' (the Reduction Principle).

\bigskip

\emph{Reduction Principle}

1. \ When a measurement of the physical property associated with an
observable $A$ is made on a system $S$ in a state $\left\vert \psi
\right\rangle $, the result only can be one of the eigenvalues of $A$.

2. The probability to get the non-degenerated value $\alpha _{i}$ is $%
P(\alpha _{i})=\left| \left\langle v_{i}|\psi \right\rangle \right| ^{2}.$

3. If the eigenvalue is degenerated then the probability is the sum over the
eigenvectors associated with this eigenvalue: $P(\alpha _{i})=\sum
\left\vert \left\langle \nu _{i}^{j}|\psi \right\rangle \right\vert ^{2}.$

4. If the measurement of $A$ on a system $S$ in the state\ $\left| \psi
\right\rangle $ has given the result $\alpha _{i}$ then the state of the
system immediately after the measurement is the normalized projection of $%
\left| \psi \right\rangle $ onto the eigensubspace of the relevant Hilbert
space associated with $\alpha _{i}$. If the eigenvalue is not degenerated
then the state of the system after the measurement is the normalized
eigenvector associated with the eigenvalue.

\bigskip

If two observables $A$ and $B$ commute then it is possible to measure both
simultaneously: the measurement of $A$ is not altered by the measurement of $%
B.$ This means that measuring $B$ after measuring $A$ does not change the
value obtained for $A$. If we again measure $A$ after a measurement of $B,$
we again get the same value for $A$. Both observables can have a definite
value.

\subsubsection{Interferences}

The archetypal example of interferences in quantum mechanics is given by the
famous two-slits experiment.\footnote{%
See, e.g., Feynman (1965) for a very clear presentation.} A parallel beam of
photons falls on a diaphragm with two parallel slits and strikes a
photographic plate. A typical interference pattern showing alternate bright
and dark rays can be seen. If one slit is shut then the previous figure
becomes a bright line in front of the open slit. This is perfectly
understandable if we consider photons as waves, as it the assumption is in
classical electromagnetism. The explanation is based on the fact that when
both slits are open, one part of the beam goes through one slit and the
other part through the other slit. Then, when the two beams join on the
plate, they interfere constructively (giving bright rays) or destructively
(giving dark ones), depending on the difference in length of the paths they
have followed. But a difficulty arises if photons are considered as
particles, as can be the case in quantum mechanics. Indeed, it is possible
to decrease the intensity of the beam so as to have only one photon
travelling at a time. In this case, if we observe the slits in order to
detect when a photon passes through (for example, by installing a
photodetector in front of the slits), it is possible to see that each photon
goes through only one slit.\ It is never the case that a photon splits to go
through both slits. The photons behave like particles. Actually, the same
experiment was done with electrons instead of photons, with the same result.
If we do the experiment this way with electrons (observing which slit the
electrons go through, i.e., sending light through each slit to
\textquotedblleft see\textquotedblright\ the electrons), we see that each
electron goes through just one slit and, in this case, we get no
interference. If we repeat the same experiment without observing which slit
the electrons pass through then we recover the interference pattern. Thus,
the simple fact that we observe which slit the electron goes through
destroys the interference pattern (two single slit patterns are observed).
The quantum explanation is based on the assumption that when we don't
observe through which slit the electron has gone then its state is a
superposition of both states \textquotedblleft gone through slit
1\textquotedblright\ and \textquotedblleft gone through slit
2\textquotedblright \footnote{%
This doesn't mean that the photon actually went through both slits.\ This
state simply can't be interpretated from a classical point of view.}, while
when we observe it, its state collapses onto one of these states. In the
first case, the position measurement is made on electrons in the superposed
state and gives an interference pattern since both states are manifested in
the measurement. In the second case, the position is measured on electrons
in a definite state and no interference arises. In other words, when only
slit 1 is open we get a spectrum, say $S_{1}$ (and $S_{2\text{ }}$when only
slit 2 is open).\ We expect to get a spectrum $S_{12}$ that sums of the two
previous spectra when both slits are open, but this is not the case: $%
S_{12}\neq S_{1}+S_{2}.$

\pagebreak

\bigskip


\begin{thebibliography}{99}
\bibitem{akerlov82} Akerlof G. A. and T. Dickens (1982) \textquotedblleft
The Economic Consequences of Cognitive Dissonance\textquotedblright , 
\textit{The American Economic Review} 72, 307-319.

\bibitem{ariely03} Ariely D. , G. Prelec and D. Lowenstein "Coherent
Arbitrariness": Stable Demand Curve without Stable Preferences. \textit{%
Quarterly Journal of Economics}, Feb. 2003, 73-103.

\bibitem{Arrow59} Arrow K. J. (1959) " Rational Choice Functions and
Orderings" \textit{Economica}, New Series, Vol 26/102, pp. 121-127.

\bibitem{Belcas81} Beltrametti E. G. and G. Cassinelli (1981), The Logic of
Quantum Mechanics, Encyclopia of Mathematics and its Applications vol. 15,
Addison-Wesley Publishing Company.

\bibitem{Benatir02} Benabou R. and J. Tirole (2002), \textquotedblleft
Self-Knowledge and Personal Motivation\textquotedblright\ \textit{Quarterly
Journal of Economics}, 117, 871-915.

\bibitem{berthoz} Berthoz A. (2003) La Decision, Odile Jacob, Paris.

\bibitem{birkhoff36} Birkhoff G. and J. von Neuman (1936) \textquotedblleft
The Logic of Quantum Mechanics\textquotedblright\ \textit{Ann. of Math}. 37,
\_823-843.

\bibitem{danlam05} Danilov V. I and A. Lambert-Mogiliansky " Non-Classical
Measurement Theory - A Framework for Behavioral Sciences" Arxiv \
xxx.lanl.gov/physics/060451

\bibitem{Eisert} Eisert J., M. Wilkens and M. Lewenstein (1999),
\textquotedblleft Quantum Games and Quantum Strategies\textquotedblright\ 
\textit{Phys. Rev. Lett. }83, 3077.

\bibitem{Erev93} Erev I., G. Bornstein and T. Wallsten (1993)
\textquotedblleft The Negative Effect of Probability Assessment on Decision
Quality\textquotedblright , \textit{Organizational Behavior and Human
Decision Processes }55, 78-94.

\bibitem{} Cohen-Tannoudji C., B. Diu and F. Laloe (1973), Mecanique
Quantique 1, Herman Editeur des Sciences et des Arts, Paris.

\bibitem{Cohen89} Cohen D. W. (1989) An Introduction to Hilbert Space and
Quantum Logic, Problem Books in Mathematics, Springer-Verlag, New York.

\bibitem{Belcas81} Beltrametti E. G. and G. Cassinelli, (1981), The Logic of
Quantum Mechanics, Encyclopedia of Mathematics and its Applications, Ed. G-C
Rota, Vol. 15, Addison-Wesley, Massachussets.

\bibitem{ben-horin79} Ben-Horin D., (1979) \textquotedblleft Dying for Work:
Occupational Cynicism Plagues Chemical Workers\textquotedblright\ in \textit{%
These Times,} June 27/July 3, 3-24.

\bibitem{feynamn } Feynman R., (1980) La Nature de la Physique, Seuil, Paris.

\bibitem{festinger57} Festinger L., (1957) Theory of Cognitive Dissonance,
Stanford University Press, Stanford, CA.

\bibitem{holland95} Holland S. S. JR. (1995) \textquotedblleft
Orthomodularity in Infinite Dimensions; a Theorem of M.
Soler\textquotedblright\ \textit{Bulletin of the American Mathematical
Society} 32, 205-234.

\bibitem{Katver00} Kahneman D. and A. Tversky (2000) Choice, Values and
Frames, Cambridge Universtity Press, Cambridge.

\bibitem{KnezCam00} Knez M. and C. Camerer (2000) \textquotedblleft
Increasing Cooperation in Prisoner's Dilemmas by Establishing a Precedent of
Efficiency in Coordination Game\textquotedblright , \textit{Organizational
Behavior an Human Decision Processes} 82, 194-216.

\bibitem{Low05} Lowenstein G. (2005) "Hot-Cold Empathy Gaps and Medical
Decision-making" \textit{Health Psychology} 24/4 549-556.

\bibitem{Gul 01} Gul F. and W. Pesendorfer (2001) `Temptation and
Self-control' \textit{Econometrica,\ }69, 1403-1436.

\bibitem{macfadden} McFadden D. (1999) \textquotedblleft Rationality of
Economists\textquotedblright\ \textit{Journal of Risk and Uncertainty} 19,
73-105

\bibitem{McFadden00} Mc Fadden D. (2000) \textquotedblleft Economic
Choices\textquotedblright , Nobel Lecture.

\bibitem{mackey63} Mackey G. W. (2004, first ed. 1963) Mathematical
Foundations of Quantum Mechanics, Dover Publication, Mineola New York.

\bibitem{Mas-colell95} Mas-Colell A., M. D. Whinston and J. R. Green (1995),
Microeconomic Theory, New York Oxford, Oxford University Press.

\bibitem{Popper} Popper K. (1992), Un Univers de Propensions,\ Edition de
l'Eclat, Paris.

\bibitem{pruitt70} Pruitt D.G.\ (1970) \textquotedblleft Reward Structure of
Cooperation: the Decomposed Prisoner's Dilemma Game\textquotedblright\ 
\textit{Journal of Personality and Psychology} 7, 21-27.

\bibitem{savage54} Savage L. J. (1954) The Foundation of Statistics, Wiley,
New York.

\bibitem{Selten1} Selten R.(1998) \textquotedblleft Features of
Experimentally Observed Bounded Rationality\textquotedblright\ \textit{%
European Economic Review} 42/3-5, 413-436.

\bibitem{tversky} Tversky A. and I. Simonson (1993) \textquotedblleft
Context-Dependent Preferences\textquotedblright\ \textit{Management Sciences 
}39, 85-117.

\bibitem{sen97} Sen A. (1997) ``Maximization and the Act of Choice''\ 
\textit{Econometrica} 65, 745-779.

\pagebreak \pagebreak
\end{thebibliography}
\end{document}